\documentclass[prb,superscriptaddress,floatfix,twocolumn]{revtex4}
\usepackage{graphicx}

\def\lsco{La$_{2-x}$Sr$_x$CuO$_4$}

\def\lbcoate{La$_{1.875}$Ba$_{0.125}$CuO$_4$}

\def\ybco{YBa$_2$Cu$_3$O$_{6+x}$}

\begin{document}

\title{Reconsidering the interpretation of quantum oscillation experiments on underdoped \ybco}
\author{J. M. Tranquada}
\affiliation{Condensed Matter Physics \&\ Materials Science Department,\\ Brookhaven National Laboratory, Upton, NY 11973-5000}
\author{D. N. Basov}
\author{A. D. LaForge}
\author{A. A. Schafgans}
\affiliation{Department of Physics, University of California, San Diego, La Jolla, California 92093}
\date{\today}
\begin{abstract}
On the basis of negative transport coefficients, it has been argued that the quantum oscillations observed in underdoped \ybco\ in high magnetic fields must be due to antinodal electron pockets.  We point out a counter example in which electron-like transport in a hole-doped cuprate is associated with Fermi-arc states.  We also present evidence that the antinodal gap in YBa$_2$Cu$_3$O$_{6.67}$ is robust to modest applied magnetic fields.  We suggest that these observations should be taken into account when interpreting the results of the quantum oscillation experiments.
\end{abstract}
\maketitle

The recent observations of quantum oscillations in underdoped \ybco\ (Refs.~\onlinecite{doir07,lebo07,seba08,jaud08}) and YBa$_2$Cu$_4$O$_8$ (Refs.~\onlinecite{yell08,bang08}) have generated considerable interest.  The oscillations have been seen in the longitudinal and Hall resistivities,\cite{doir07,lebo07,bang08} as well as the magnetization, \cite{seba08,jaud08,yell08} as a function of magnetic field at very low temperature.   While the Hall resistivity is positive in the normal state, it is negative when the oscillations are observed.  The consensus interpretation is that the experiments imply the presence of electron-like pockets at the Fermi surface, under the experimental conditions of high magnetic field ($\agt40$ T) and low temperature ($T<5$ K).  The evidence for small electron pockets came as a considerable surprise, as angle-resolved photoemission spectroscopy (ARPES) studies in the normal state of underdoped cuprates show no evidence for such pockets.  Instead, one generally observes a gapless Fermi arc and a large pseudogap in the ``antinodal'' region of reciprocal space.\cite{dama03,norm05,hoss08,naka09} The common theoretical approach to this problem has been to invoke some sort of competing density-wave order that causes a reconstruction of the Fermi surface, leading to small electron pockets in the ``antinodal'' region of reciprocal space.\cite{mill07,harr07,chak08b,chak08,kuch08,chen09,sent09,tail09,dimo08,podo08}  However, this leaves us with a new problem: how does one reconcile antinodal pockets with the large pseudogap observed in the zero-field normal state?  This conundrum has motivated further theoretical work.\cite{jia09,sent09}

The purpose of this paper is to present two phenomenological observations that bear on the interpretation of the quantum oscillation experiments.  The first has to do with interpretation of the Fermi-surface pockets as being electron like: in the one known case of a hole-doped cuprate where a crossover to electron-like transport is observed in zero magnetic field (thus enabling ARPES measurements), the behavior appears to be associated with the Fermi arc states.  The second observation has to do with the likelihood of antinodal Fermi-surface pockets: optical conductivity measurements on \ybco\ indicate that the antinodal gap is fairly robust in a magnetic field.  These results suggest that a reconsideration of the theoretical interpretation of the quantum oscillation experiments may be in order.

\begin{figure}[b]
\centerline{\includegraphics[width=2.5in]{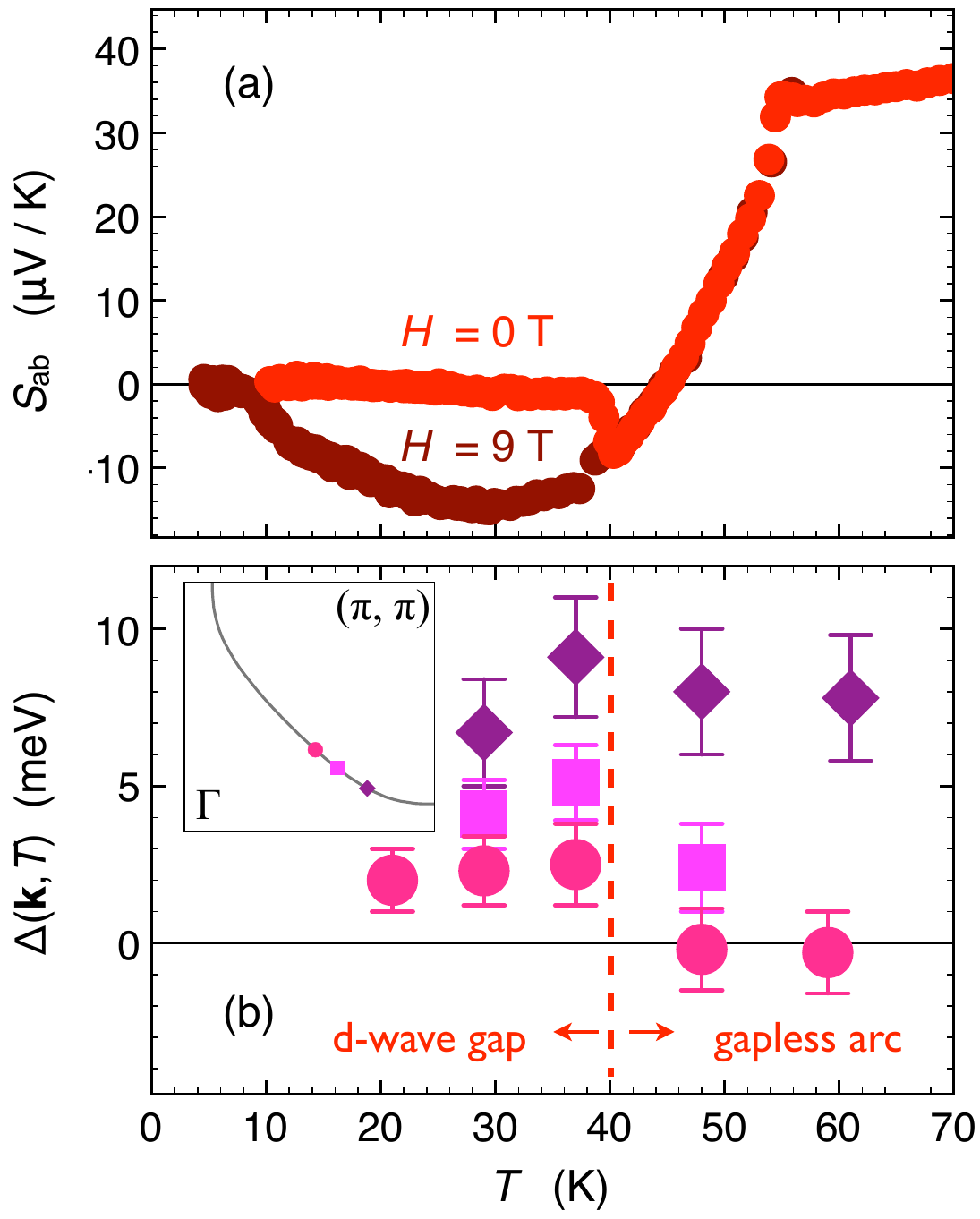}}
\caption{(color online) Collected results for \lbcoate: (a) In-plane thermopower, $S_{ab}$, from Ref.~\onlinecite{li07}.  (b)  Gap in the electronic spectral function at several {\bf k} points, indicated in the inset, from the ARPES study of He {\it et al.}\cite{he09}  
}
\label{fg:lbco}
\end{figure}

The case of electron-like transport occurs in \lbcoate\ (and also\cite{huck98,chan09} in rare-earth-doped \lsco\ with $x\sim\frac18$).  Figure~\ref{fg:lbco}(a) shows the in-plane thermoelectric power, $S_{\rm ab}$, taken from Ref.~\onlinecite{li07}.  While $S_{\rm ab}$ has a substantial positive value above 54~K, it drops rapidly below that temperature, becoming negative below 45 K.  Below 40~K, the thermpower approaches zero due to superconducting correlations; when the superconductivity is suppressed with a magnetic field, the thermopower remains negative.  The sign of the thermopower, like the sign of the Hall coefficient, is generally interpreted (in a one band model) as the sign of the carriers.  Based on such an interpretation, we appear to have electron-like carriers for $T<45$~K.

For comparison, Fig.~\ref{fg:lbco}(b) shows the temperature dependence of the energy gap measured at several points on the nominal Fermi surface in \lbcoate\ by He {\it et al.}\cite{he09}  For $T\agt 40$~K, including the region where $S_{\rm ab}<0$,  there is a gapless Fermi arc.  Below 40~K, where $S_{\rm ab} \approx 0$, there is a $d$-wave-like gap on the Fermi arc.  We conclude that the negative thermopower must be associated with the Fermi arc states.  Note that the antinodal states remain gapped throughout this temperature range.

In a recent study, Chang {\it et al.}\cite{chan09} have shown that the temperature dependence of the thermopower in YBa$_2$Cu$_3$O$_{6.67}$, measured in a high magnetic field, is very similar to that found in \lbcoate.  In particular, it goes negative for $T\alt 50$~K, and a field of 8 T is sufficient to access this negative thermopower state down to $\sim30$~K.  (For a similar temperature-dependent sign change in the Hall coefficient, see Fig.~S1b in the supplementary material to Ref.~\onlinecite{lebo07}.)
It would be very surprising if the physics behind the negative thermopower in these different systems were not the same.  (We note that a new paper\cite{sing09} reports quantum oscillations in \ybco\ with $x$ as large as 0.69.)

It is also of interest to consider probes that are directly sensitive to the development of coherent antinodal pockets induced by magnetic field at low temperature.  One such probe is optical conductivity measured with the polarization of the light along the $c$ axis.  The antinodal states are crucial to conduction between the planes,\cite{chak93} and $c$-axis optical conductivity measurements on underdoped \ybco\ provided one of the first sightings of the electronic pseudogap.\cite{home93b}   

\begin{figure}[t]
\centerline{\includegraphics[width=3.4in]{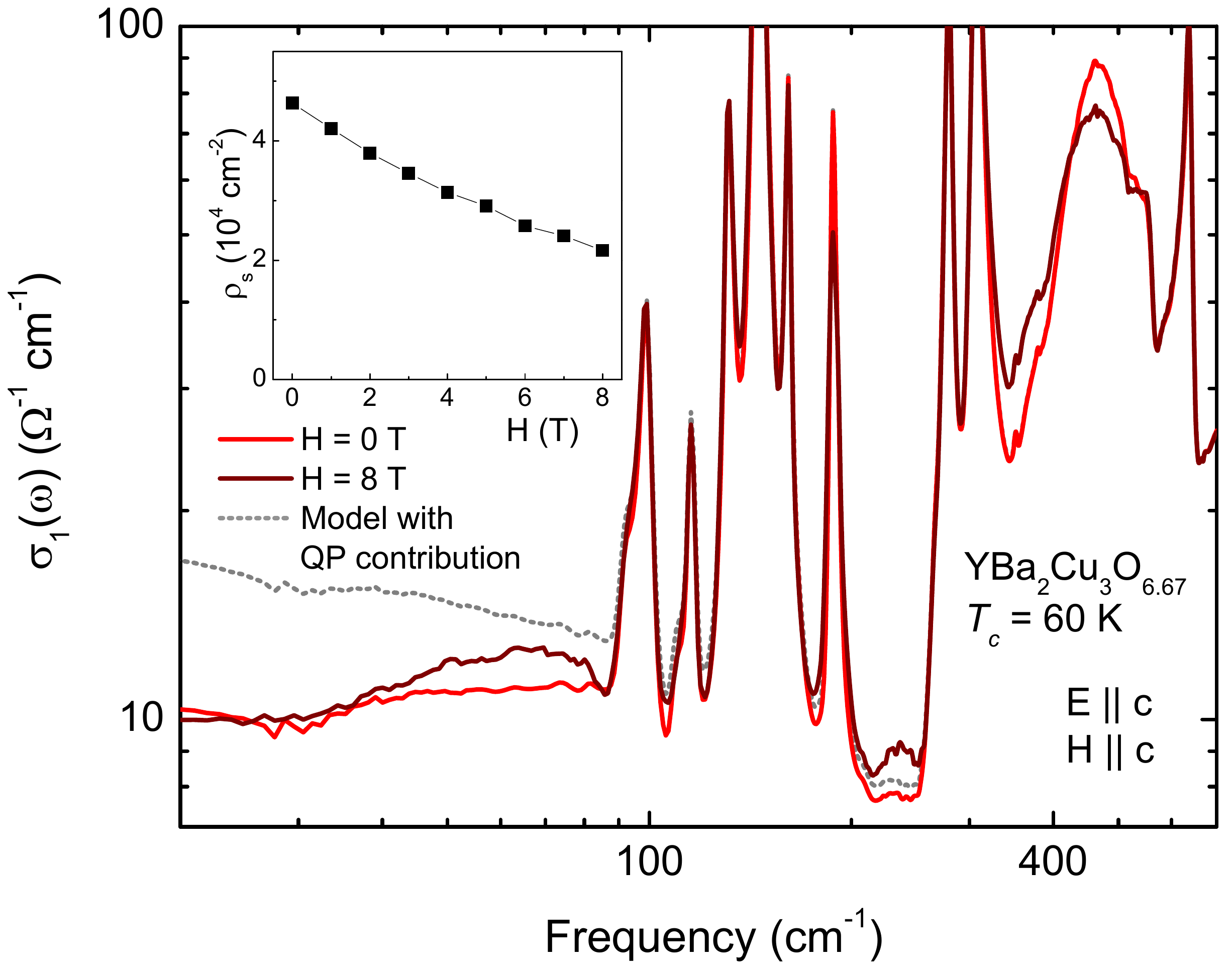}}
\caption{(color online) The interlayer $c$-axis conductivity for YBa$_2$Cu$_3$O$_{6.67}$ at 8K in zero magnetic field and at 8 T, using results from Ref.~\onlinecite{lafo08}.  Inset: the magnetic field dependence of the superfluid density $\rho_s(H)$. Dashed (gray) line, main panel: a hypothetical result assuming that the oscillator strength corresponding to $\rho_s(0T) - \rho_s(8T)$ is transferred to a coherent $c$-axis response (Drude peak), with scattering rate $1/\tau=50$~cm$^{-1}$.}
\label{fg:opt}
\end{figure}

For the real part of the $c$-axis conductivity, $\sigma_{1,c}(\omega)$, the pseudogap shows up in the normal state as a strong suppression of the low-frequency electronic conductivity as the temperature is reduced towards the superconducting transition temperature, $T_c$.  For YBa$_2$Cu$_3$O$_{6.67}$ ($T_c=60$~K), recent measurements\cite{lafo09} show that the low-frequency limit of $\sigma_{1,c}$  is approximately 25~$\Omega^{-1}$~cm$^{-1}$ at room temperature; in contrast, the value at $T_c$ is already very close to that at $T<<T_c$, where it reaches 10~$\Omega^{-1}$~cm$^{-1}$.  If a strong magnetic field suppressed the pseudogap, or induced coherent states at the Fermi level, we would expect to see an increase in the low-frequency conductivity.

The impact of  $c$-axis magnetic fields (up to 8 T) on $\sigma_{1,c}(\omega)$ for YBa$_2$Cu$_3$O$_{6.67}$  has recently been reported.\cite{lafo08,lafo09}   In Fig.~\ref{fg:opt}, we show the low-temperature results ($T=8$~K) on log-log scales to emphasize the low-frequency and low-conductivity regimes.  Application of a magnetic field of 8~T causes remarkably little change in $\sigma_{1,c}(\omega)$, even though this field is sufficient to reduce the superfluid density by 50\% (see inset of Fig.~\ref{fg:opt}).  If the electron density removed from the superfluid were transferred to coherent antinodal quasiparticles, then we would expect to regain a substantial fraction of the low-frequency conductivity found at room temperature.  An estimate of that response is indicated by the dashed line in Fig.~\ref{fg:opt}.  In actuality, the spectral weight removed from the condensate in finite fields is transferred to frequencies above 1000 cm$^{-1}$ ({\it i.e.}, $\agt100$ meV), which is above the pseudogap energy.\cite{lafo08} The absence of any significant field-induced spectral weight at low frequencies indicates that the antinodal gap is rather robust. 

The maximum field used in the optical conductivity experiment is certainly much less than the threshold field for observing quantum oscillations ($\sim30$~T).  Nevertheless, as already pointed out, Chang {\it et al.}\cite{chan09} have shown for this composition\cite{note3} that 8~T is sufficient to access the phase with electron-like transport at higher temperatures.  At 8~K, where the optical measurements were done, we expect that the ``normal'' state associated with the magnetic vortex cores corresponds to this same phase with electron-like transport.  The absence of any field-induced Drude component in $\sigma_{1,c}(\omega)$ suggests to us that there is no coherent single-particle weight at the Fermi level in the antinodal region.   Note that this argument is independent of whether or not there is any competing order present.  If there are no quasiparticles at the Fermi level in the antinodal region, then it does not matter whether there might be nominal pockets in that region due to Fermi-surface reconstruction.  One possibility that we cannot rule out is that the quasiparticle weight is finite but incredibly small at 8~T.  In that case, the weight would certainly be larger at the much higher fields of the quantum oscillation regime.  Further analysis and discussion of this problem are presented in Ref.~\onlinecite{lafo10}. 

To summarize, we have made two phenomenological observations that are relevant to the interpretation of quantum oscillation experiments on underdoped cuprate superconductors.  In the first case, we have pointed out an example in which electron-like transport behavior is associated with Fermi-arc states.  While we cannot explain why this occurs, we believe that the experimental facts are solid.  In the second case, we have shown that $c$-axis optical conductivity measurements provide direct evidence that moderate magnetic fields have negligible impact on the antinodal pseudogap.  We suggest that these are good reasons to question common assumptions and to reconsider the theoretical interpretation that the quantum oscillation experiments must be due to antinodal electron pockets.


We are grateful to S. A. Kivelson, C. Panagopoulos, and O. Vafek for helpful discussions. Work at Brookhaven was supported by the Office of Science, U.S. Department of Energy under Contract No.\ DE-AC02-98CH10886. Work at UCSD is supported by NSF DMR 0705171.

{\it Note added}:  A new theory paper has proposed that quantum oscillations could result from Fermi arcs combined with antinodal pairing gaps.\cite{pere10}  Those authors did not consider the effective sign of the charge carriers.  Combining their idea with our phenomenological observation of electron-like response associated with Fermi arc states in the stripe-ordered phase provides a possible alternative to the antinodal pocket scenario for interpreting quantum oscillation experiments.


\end{document}